\documentclass[aip,cha,reprint]{revtex4-1}

\usepackage[colorlinks=true,linkcolor=blue,citecolor=blue,urlcolor=blue,breaklinks=true]{hyperref}
\usepackage{graphicx}
\usepackage{amsfonts}
\usepackage{amsmath}
\usepackage{color}

\newcommand{\eps}{\varepsilon}

\newcommand{\basin}[2]{\mathcal{B}_{#1}(#2)}

\newcommand{\erfc}{\text{erfc}}

\newcommand{\pth}{p_\text{th}}

\newcommand{\TrapInt}{I_\text{trap}}

\begin{document}

\title{Controlling systems that drift through a tipping point}

\author{Takashi Nishikawa} 
\email{t-nishikawa@northwestern.edu}
\affiliation{Department of Physics \& Astronomy, Northwestern University, Evanston, IL 60208, USA}
\author{Edward Ott}
\affiliation{Institute for Research in Electronics and Applied Physics, University of Maryland, College Park, MD 20742, USA}

\begin{abstract}
Slow parameter drift is common in many systems (e.g., the amount of greenhouse gases in the terrestrial atmosphere is increasing).
In such situations, the attractor on which the system trajectory lies can be destroyed, and the trajectory will then go to another attractor of the system.
We consider the case where there are more than one of these possible final attractors, and we ask whether we can control the outcome (i.e., the attractor that ultimately captures the trajectory) using only \textit{small} controlling perturbations.
Specifically, we
consider the problem of controlling a noisy system whose parameter slowly drifts through a saddle-node bifurcation taking place on a fractal boundary between the basins of multiple 
attractors.
We show that, when the noise level is low, a small perturbation of size comparable to the noise amplitude applied at a single point in time can ensure that the system will evolve toward a target attracting state with high probability.
For a range of noise levels, we find that the minimum size of perturbation required for control is much smaller within a time period that starts some time after the bifurcation, providing a ``window of opportunity'' for driving the system toward a desirable state.
We refer to this procedure as \textit{tipping point control}.
\end{abstract} 

\onecolumngrid
\noindent
{\hfill\small \href{\doibase 10.1063/1.4887275}{Chaos \textbf{24}, 033107 (2014)}}

\maketitle

\begin{quotation}

There is a growing concern that the observed slow increase of greenhouse gases may lead to a sudden, dramatic change in the global climate system, profoundly impacting our society.
This is but one (rather striking) example of what has been called a ``tipping point.''
From the dynamical systems point of view the general tipping point scenario can be modeled as a system bifurcation with a slowly varying parameter.
In this paper, 
we demonstrate that, if such a system has multiple post-bifurcation attractors, then it may be controlled to a desirable state by a small, carefully chosen perturbation, applied only once at an appropriate timing. 
We call this ``tipping point control.''
The general mathematical framework we employ suggests potential application not only to climate change, but to other examples, such as a power grid experiencing slow increase in demand and a food web having a gradually decreasing population of a certain species due to overhunting.

\end{quotation}

\section{Introduction}

Physical systems with a slowly varying parameter is common in real world; take for example a climate system experiencing gradual increase in 
greenhouse gases,
or a power grid subject to slow rise in demand.
When variation of a parameter 
adiabatically
pushes such a system through bifurcation and causes a loss of the stability of the current 
attracting
state (popularly often referred to as a ``tipping point''),
predicting and possibly controlling the future evolution of the system is a problem of critical concern, which is closely related to the field of dynamical bifurcation\ \cite{Benoit:1990qy,Berglund:2006mz,Ashwin:2012fk,PhysRevE.85.046202}.
This could
be particularly important if multiple stable states exist after the bifurcation, and if one of them corresponds to a catastrophic system-wide event, such as an abrupt drop in global mean temperature in a climate model, signifying a transition into an ice age (see Ref.\ \onlinecite{Alley:2003fj} for review on this and other abrupt climate changes), 
or
a voltage collapse in power systems\ \cite{54571,136788}, which can cause a large-scale power outage.
Furthermore, for a large system (e.g., the Earth's climate), in order to be feasible, one would like the control to be accomplished using relatively small perturbations.
We refer to this type of control as ``tipping point control.''
We emphasize that application of the general tipping point control method proposed and illustrated in this paper requires knowledge of an accurate system model, which in some cases of interest (e.g., climate) is not currently known (but could become known in the future).

For some types of bifurcation, there will be a natural stable state near the pre-bifurcation state, which the system will robustly follow after the bifurcation.
For others, 
like the saddle-node bifurcation we consider here, there will be no such state in the vicinity of the bifurcation point, and the future course of the system trajectory will depend on the global structure of the basins of attraction for the post-bifurcation attractors.
Here a basin for an attractor at a given time is defined as the set of all states from which the system will evolve to the attractor. 
The structure of basin boundaries slowly varies with time because of the drifting parameter.
If the boundary is fractal, then the bifurcation can be indeterminate, in the sense that the fate of the system after the bifurcation 
(the final attractor)
can depend on 
small noise or be extremely sensitive to the specific rate of parameter variation \cite{thompson1991indeterminate,soliman1992indeterminate,thompson1992global,soliman1995dynamic,Breban:2003fk,PhysRevE.68.066213,THOMPSON:2011qy}.
Thus, prediction of the final attractor can be difficult.
The flip side of this is 
that, in a well-measured and well-characterized system,
relatively
small change in the system state has the potential to change the course of system evolution 
dramatically.
This is similar to the fact that sensitive dependence on initial conditions, which is a defining characteristic of chaotic systems, allows one to control a noiseless chaotic system with arbitrarily small perturbation\ \cite{Shinbrot:1993yq,Ott:2002fk,PhysRevLett.64.1196,Schroer:1997uq,PhysRevLett.68.2863,Bollt:1995kx,PhysRevE.51.102,Steingrube:2010fk}.
When noise is present in the parameter-drifting system, there will be a minimum size of perturbation required for control, and the dependence of this minimum on the noise level and the timing of the control is the subject of this study.

Specifically, we show that the minimum size of a single perturbation required for effective control at a specific time is comparable to the noise amplitude.
Moreover, for a range of noise levels, the time at which the perturbation is applied appears to matter---there is a period of time in which the minimum perturbation size is much smaller---and, curiously, this preferred time window starts some time \emph{after} the parameter value passes the bifurcation point.
Analogous dependence of control effectiveness on the timing has recently been found in the context of controlling networks\cite{Rio}.
Besides numerical verification for a specific one-dimensional map, we provide a 
general argument that explains this behavior.

\section{Illustration of tipping point control}
Consider a one-parameter family of one-dimensional maps, $x_{n+1} = f_\mu(x_n)$, $n=0,1,\ldots$, which has a backward saddle-node bifurcation at $\mu = \mu_*$.  
This means that a pair of attracting and repelling fixed points, which exist when $\mu < \mu_*$, coincide at $x = x_*$ when $\mu = \mu_*$ and disappear when $\mu > \mu_*$.
Suppose that both before and after the bifurcation (i.e., 
for $\mu \in (\mu_* - \eps, \mu_* + \eps)$ for some $\eps > 0$),
there are at least two other attractors for the system whose basins share a fractal boundary.
Suppose further that the fractal boundary contains the saddle-node bifurcation point $x = x_*$ when $\mu = \mu_*$.
This situation occurs when a fixed-point attractor, located in a basin having the so-called Wada property, disappears by colliding with a saddle on the boundary through a saddle-node bifurcation~\cite{PhysRevLett.75.2482}.
Ref.\,\onlinecite{Kennedy1991213} argues that Wada basins are quite common.

Now consider a slow variation of the parameter $\mu$ from $\mu_s$ to $\mu_f$ (assuming $\mu_* - \eps < \mu_s < \mu_* < \mu_f < \mu_* + \eps$) at a rate of $\delta\mu \ll 1$ per iterate:
\begin{equation}\begin{split}\label{system}
x_{n+1} &= f_{\mu_n}(x_n) + A \xi_n,\\
\mu_{n+1} &= \mu_n + \delta\mu,
\end{split}\end{equation}
where $\mu_0 = \mu_s$ and $\xi_n$ is white noise of unit amplitude, 
$\langle \xi_n \xi_{n'} \rangle = \delta_{nn'}$ (where $\langle\cdots\rangle$ denotes the expectation value  of $\cdots$),
so that $A$ represents the noise intensity.
To facilitate our numerical experiments, we stop varying $\mu$ as soon as $\mu \ge \mu_f$ (we define $n_f$ to be the corresponding value of $n$) and determine which attractor the system approaches by iterating the map $f_{\mu}$ further with the fixed $\mu=\mu_{n_f}$ for $n \ge n_f$, as in Refs.\ \onlinecite{PhysRevE.68.066213,Breban:2003fk}.

We consider the following control problem: given the current state of a system that is destined to evolve into an undesirable state, can we apply a small perturbation to the current state, so that the system will go to a desirable state with high probability?
If such a control is possible, how large does the perturbation need to be?
Formally, given a 
realization of a noisy
trajectory $\{ x_n \}_{n=0}^\infty$ of the system~\eqref{system} that converges to an undesirable attractor $A_u$, we apply a small perturbation $x'_{n_c} = x_{n_c} + \delta x$ at the $n_c$-th iterate.
Denote by $\{ x'_n \}_{n=0}^\infty$ the new trajectory defined by $x'_n = x_n$ for $n<n_c$ and by the recursive application of Eq.\ \eqref{system} for $n > n_c$.
We then seek the minimum size $|\delta x|$ of all such perturbations for which $\{ x_n \}_{n=0}^\infty$ converges to the desirable attractor $A_d$ with probability larger than a given threshold $\pth$.

Note that applying a perturbation before the bifurcation point $\mu_*$ is not effective since the bifurcation is indeterminate.
The fine-scale structure of the fractal basin boundary near the bifurcation point implies that, even when the system is perturbed into the basin of the desirable attractor before the bifurcation, arbitrarily small noise can nudge the system into the basin of an undesirable attractor as it passes through the neighborhood of the bifurcation point.
Thus, one needs to wait at least until the parameter $\mu$ exceeds $\mu_*$ to apply an effective control.
We will see, however, that it is actually better to wait a bit longer.

As an illustrative example, consider the system\ \eqref{system} with $f_\mu$ given by 
\begin{equation}\label{BNO_map}
f_\mu(x) = g^{[3]}(x) + \mu\sin(3\pi x),
\end{equation}
where $g^{[3]}$ is the third iterate of the logistic map, $g(x) = 3.832x(1-x)$.
This map satisfies all the assumptions made above, and a saddle-node bifurcation occurs at $x_* \approx 0.15970$ with $\mu_* \approx 0.00279$.
We set the range of parameter sweep to be from $\mu_s = 0$ to $\mu_f = 0.0045$ with a rate of $\delta\mu = 2 \times 10^{-5}$.
For the desirable attractor $A_d$, we choose the attracting fixed point for $f_{\mu_f}$ located at $x \approx 0.49585$, and the only other attractor for $f_{\mu_f}$ is the fixed point near $x \approx 0.95919$, which will be the undesirable attractor $A_u$.
The scaling properties of this system were studied in Ref.\ \onlinecite{PhysRevE.68.066213}.

Note that, while, for simplicity, we consider the control to be a state change $\delta x$ at a given time, other controls could also be considered.
Examples include a one-time change in $\mu$ or some other supposed system parameter, a change that operates over several time steps, etc.
Although such different controls would change details of what happens in a given numerical experiment, we do not expect our general conclusions to change.

\section{Results}

\begin{figure}
\begin{center}
\includegraphics[width=3.4in]{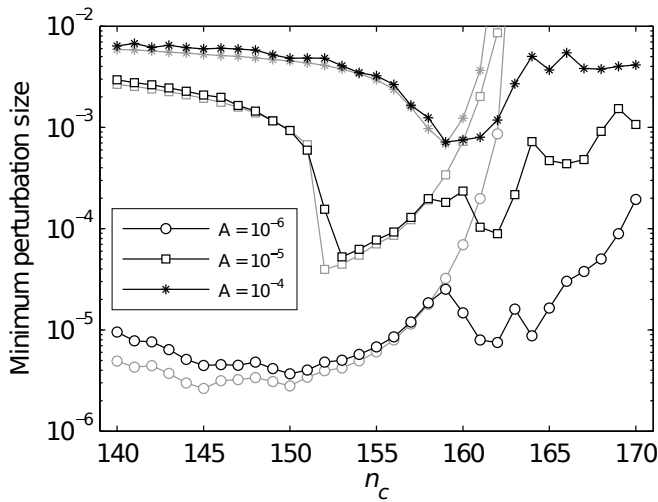}
\end{center}
\caption{\label{figure_min_kick_size}
The black curves show the minimum
size of perturbation 
$|\delta x|$
required for successful control along typical trajectories just passed the bifurcation point.
The parameter $\mu$ passes the bifurcation value $\mu_*$ for the first time at $n = n_* = 140$.
For a given noise amplitude $A$ and a point $n=n_c$ on a reference trajectory $\{ x_n \}_{n=0}^\infty$, the minimum value of the perturbation size $|\delta x|$ was estimated by incrementing $|\delta x|$ until the success probability $p(x_{n_c}, n_c)$ is larger than $\pth = 0.9$ using a variable increment size, which is initially equal to $10^{-3}$ and adaptively made smaller to ensure a relative resolution of $0.005$ with respect to the computed $|\delta x|$.
The estimated minimum was averaged over 10 realizations of a noisy reference trajectory starting from the point $x_0 = x_*$.
The gray plots are the corresponding theoretical prediction based on the probability estimate for basin intervals provided by Eq.~\eqref{prob_func_est}.}
\end{figure}

To investigate the controllability of this system, we perform the following experiment.
For a given noise amplitude $A$, we generate an ensemble of noisy reference trajectories of the system\ \eqref{system} that converge to the undesirable attractor $A_u$, which represents typical paths of the system slowly undergoing the bifurcation and eventually settling into an undesirable state.
The initial condition $x_0$ is chosen to be the bifurcation point $x_*$.
Given $\delta x$ and $n_c$, let $p(x'_{n_c}, n_c)$ denote the probability that the perturbed state $x'_{n_c} = x_{n_c} + \delta x$ evolves to the desirable attractor $A_d$ under the influence of noise.
We estimate the minimum value of $|\delta x|$ for which this probability exceeds the threshold $\pth = 0.9$.
The result of this simulation is shown in Fig.\ \ref{figure_min_kick_size} for a range of noise amplitudes. 

We find that the system can be controlled successfully with perturbations of size comparable to the noise amplitude, as can be seen from the minima of the curves in Fig.\ \ref{figure_min_kick_size}.
The control-to-noise ratio for these minima ranges from $3.3$ to $9.8$.
Note, however, that the required perturbation size is much smaller in a relatively narrow range of time for $A = 10^{-5}$ and $10^{-4}$.
Thus, for a range of noise levels, there appears to be a \emph{time window} of opportunity for 
effective, low-amplitude control. 
Interestingly, the curves for $A=10^{-6}$ and $10^{-5}$ also show a second minimum.

\begin{figure}
\begin{center}
\includegraphics[width=3.3in]{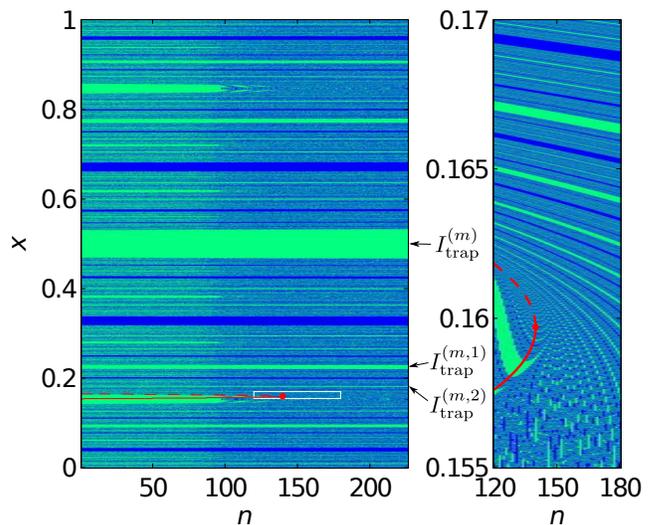}
\end{center}
\caption{\label{figure_basins}
Basins of attraction for the system\ \eqref{system} with $f_\mu$ given in Eq.\ \eqref{BNO_map} and a slowly varying parameter $\mu$ with a sweep speed of $\delta\mu = 2 \times 10^{-5}$ per iteration.
The green and blue regions indicate the basins of the desirable attractor $\basin{n}{A_d}$ and undesirable attractor $\basin{n}{A_u}$, respectively.
A pixel with an intermediate shade between the two colors is used to represent the fraction of points in the two basins in the corresponding small interval of $x$ (of width $\approx 10^{-3}$ for the left panel and $\approx 3 \times 10^{-6}$ for the right panel), estimated by a random sample of $20$ points in the interval.
The red dashed and solid curves indicate the saddle and node, respectively, which are destroyed at the bifurcation point (red dot).
The white rectangle in the left panel indicates the region blown up in the right panel.
As indicated to the right of the left panel, some
of the green stripes in the left panel correspond to 
the basin intervals $\TrapInt^{(m)}$, $\TrapInt^{(m,1)}$, and $\TrapInt^{(m,2)}$ for $0 \le m \le n_f = 226$.
}
\end{figure}

\section{Theory}

A key idea for understanding the above result is that the system can in principle be controlled by perturbing its state into the basin of the desirable attractor $A_d$, which we denote by $\basin{n}{A_d}$.
Here the basin of attraction for $A_d$ at time $n$ is defined
for system\,\eqref{system} with arbitrary $f_{\mu}$ 
as the set of all points $x$ such that the noiseless trajectory $\{ x'_k\}_{k=n}^\infty$ with $x'_n = x$ (and $A=0$) converges to $A_d$.
For each $n \ge n_f$, the basin $\basin{n}{A_d}$  has identical fractal structure and is composed of an infinite number of intervals, since $\mu_n$, and thus the map $f_{\mu_n}$, is fixed for $n \ge n_f$.
For $0 \le n < n_f$, the basins are determined recursively by the relation  $\basin{n}{A_d} = f_{\mu_n}^{-1} \bigl(\basin{n+1}{A_d}\bigr)$, so each $\basin{n}{A_d}$ is a union of 
an
 infinite number of intervals, which is illustrated in Fig.\ \ref{figure_basins} for the specific $f_{\mu}$ given by Eq.\,\eqref{BNO_map}.
In general, the fractal nature of $\basin{n}{A_d}$, which is a direct consequence of our assumption that $f_{\mu}$ has a fractal basin boundary,
allows for controlling the system by an arbitrarily small perturbation in the absence of noise.

\begin{figure}
\begin{center}
\includegraphics[width=2.9in]{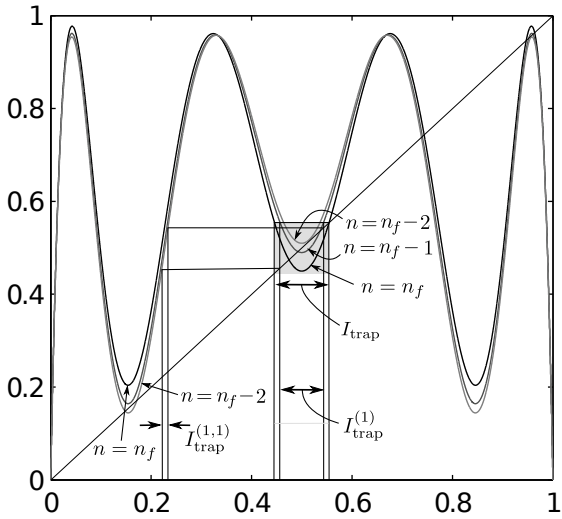}
\end{center}
\caption{\label{figure_trap_int}
Definition of trapping intervals $\TrapInt^{(m)}$ and $\TrapInt^{(m,k)}$.
The three curves represent the graphs of $f_{\mu_n}$ 
at 
three consecutive 
times
of $n = n_f-2, n_f-1$, and $n_f$.
Note that while these are accurate graphs of $f_{\mu}$ for some $\mu$, the difference between the $\mu$ values are exaggerated for clarity of presentation.
}
\end{figure}

To derive a general theory for one-dimensional system~\eqref{system} that explains the results 
shown as the black curves
in Fig.~\ref{figure_min_kick_size} 
and provides the
theoretical predictions shown as the gray curves in 
the same figure,
we define a characteristic subset of these basin intervals as follows.
First, there is a trapping interval for $f_{\mu_f}$, which has the property that every trajectory that enters $\TrapInt$ stays in the interval under iterations of $f_{\mu_f}$ [i.e., $f_{\mu_f}(\TrapInt) \subset \TrapInt$] and converges to $A_d$. 
We then define an analogous trapping interval $\TrapInt^{(m)}$ as the $m$-th pre-image of $\TrapInt$ under the system\ \eqref{system} that contains the attractor $A_d$.
Thus, any trajectory that starts in $\TrapInt^{(m)}$ at time $n = n_f - m$ will stay near $A_d$ and eventually approach $A_d$.
These intervals have approximately the same width for each $m$.
Next, we define $\TrapInt^{(m,1)}$ as the pre-image of the trapping interval $\TrapInt^{(m)}$ under the monotonically increasing segment of the map $f_{\mu_n}$ immediately to the right of the minimum point associated with the saddle-node bifurcation (see Fig.\ \ref{figure_trap_int}).
Similarly, we define $\TrapInt^{(m,k)}$ as the $k$-th pre-image of $\TrapInt^{(m)}$ using the corresponding part of the maps $f_{\mu_n}$ for $k = 1,2,\ldots$.
The basin interval $\TrapInt^{(m,k)}$ is thus characterized by the fact that a trajectory starting from it at time $n = n_f - m - k$ will visit the sequence of $k$ intervals, $\TrapInt^{(m,k-1)}, \TrapInt^{(m,k-2)}, \ldots, \TrapInt^{(m,1)}, \TrapInt^{(m,0)} (= \TrapInt^{(m)}$), stay near the attractor $A_d$ for $m$ iterations until $n=n_f$ (visiting $\TrapInt^{(m-1)},\ldots, \TrapInt^{(0)}$), and then converge to $A_d$.
Other basin intervals in $\basin{n}{A_d}$ can similarly be characterized by pre-images of the trapping intervals defined through different monotonic segments of $f_{\mu_n}$.
We remark that the definition of the trapping intervals applies to system\,\eqref{system} with arbitrary $f_{\mu}$ that satisfies the assumptions stated in the first paragraph of Sec.\,II.

\begin{figure}
\begin{center}
\includegraphics[width=3.3in]{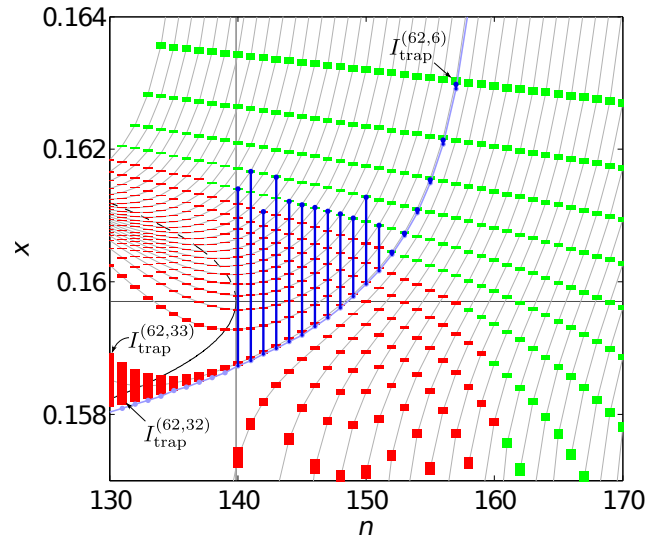}
\end{center}
\caption{\label{figure_basin_int}
Basin intervals $\TrapInt^{(m,k)}$ and minimum perturbation size for controlling noisy trajectories.
The intervals are represented by the vertical extent of the red and green boxes.
The centers of the boxes are connected by gray lines, indicating how one interval maps to another under time evolution of the system dynamics\ \eqref{system} (i.e., from $\TrapInt^{(m,k)}$ to $\TrapInt^{(m,k-1)}$).
The green (red) boxes correspond to the intervals for which the estimated probability $p(x_n, n)$ is greater than $\pth=0.9$ (less than $0.9$). 
The probabilities were estimated using Eq.\ \eqref{prob_func_est} with $A = 10^{-5}$. 
The light blue curve is one of the reference trajectories used for the $A=10^{-5}$ result in Fig.\ \ref{figure_min_kick_size}.
A blue vertical line with a dot represents a perturbation $\delta x$ of the minimum magnitude for the corresponding point on the trajectory. 
The dashed and solid black curves that meet at the intersection of vertical and horizontal gray lines are the saddle and node points, respectively, which undergo bifurcation at $x = x_*$.
}
\end{figure}

Notice that intervals are typically stretched by the action of $f_{\mu_n}$ except for those special trapping intervals $\TrapInt^{(m)}$, which contain $A_d$, whose width is approximately independent of $m$, and any subinterval of which shrinks under the iterations of the system.
This stretching is expected for the class of systems considered here, since the assumed fractal basin boundary is usually associated with (transient) chaotic dynamics, which stretches small intervals around trajectory points on average.
Thus, a given basin interval would typically be narrower if it takes a larger number of iterates to map to the basin interval that contains $A_d$.
Based on this observation, we expect that 
the widest basin intervals for any given $n$ are given by the sequence of intervals $\{ \TrapInt^{(m,k)} \,\vert\, n+k+m=n_f, \,\,k=0, \ldots, n \}$.
These intervals are shown  in Fig.\ \ref{figure_basin_int} for 
the particular case given by Eq.\,\eqref{BNO_map}.

\begin{figure}
\begin{center}
\includegraphics[width=3.3in]{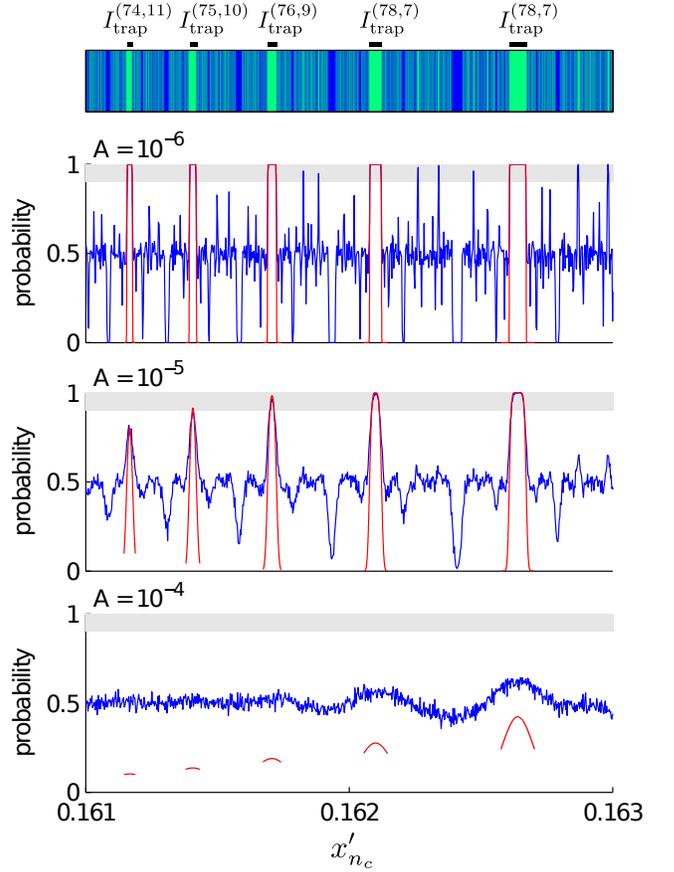}
\end{center}
\caption{\label{figure_prob_func}
Success probability $p(x'_{n_c}, n_c)$ as a function of $x'_{n_c}$ for a fixed $n_c$ and noise amplitude $A= 10^{-6}, 10^{-5}$, and $10^{-4}$.
We used $n_c=140$, which corresponds to immediately after the parameter $\mu$ exceeds the bifurcation value $\mu_*$. 
The probability function, shown as blue curves, was estimated by the fraction of trajectories that entered the trapping interval $\TrapInt$ using $400$ noise realizations.
The red curves are the theoretical prediction\ \eqref{prob_func_est} in the vicinity of the basin intervals $\TrapInt^{(m,k)}$ with $n_c + m + k = n_f = 226$ (or equivalently, $m + k = 85$), represented by small black bars at the top of the figure.
The gray horizontal bars at the top of each panel indicates the condition $p(x'_{n_c}, n_c) > \pth$.
The horizontal box at the top of the figure shows the basins, color-coded with green (desirable attractor $A_d$) and blue (undesirable attractor $A_u$).
An intermediate shade between the two colors is used to represent a mixture of the two basins in a small neighborhood of size $\approx 3 \times 10^{-6}$, corresponding to a single pixel in horizontal direction in this panel.}
\end{figure}

These special basin intervals are relevant for determining the minimum required size of 
the control perturbation $|\delta x|$, since the wider the basin intervals visited by the noisy trajectory, the more likely it is to approach the desirable attractor $A_d$.
A trajectory started from $x'_{n_c}$ in $\TrapInt^{(m,k)} \in \basin{n_c}{A_d}$ would visit the basin intervals $\TrapInt^{(m,k-1)}, \TrapInt^{(m,k-2)}, \ldots, \TrapInt^{(m,0)}(=\TrapInt^{(m)}), \TrapInt^{(m-1)}, \ldots, \TrapInt^{(1)}, \TrapInt^{(0)}(=\TrapInt), \TrapInt, \ldots$, and converge eventually to $A_d$, but it can jump outside these intervals at any time under the influence of noise. 
The probability of such an event is determined by the amplitude of noise relative to the size of the basin interval the trajectory is visiting.
For fixed $n_c$ and $A$, the probability $p(x'_{n_c}, n_c)$ that a perturbed system state $x'_{n_c} = x_{n_c} + \delta x$ evolves under time evolution to $A_d$ is a function of $x'_{n_c}$.
This is shown in Fig.\ \ref{figure_prob_func} for the case of Eq.\,\eqref{BNO_map}.
This probability was shown in Ref.\ \onlinecite{PhysRevE.68.066213} to be a function of the scaled variable $A/\delta\mu^{5/6}$ for a fixed $x'_{n_c}$.
We see that in order for $p(x'_{n_c}, n_c) > \pth$, one needs to have $x'_{n_c}$ ``deep enough'' in a basin interval, which is impossible if the interval is too narrow compared to the noise amplitude.
Thus, the minimum size of a control perturbation is determined by the minimum distance to a basin interval that is wide enough to make $p(x'_{n_c}, n_c) > \pth$ somewhere in the interval.

How wide is wide enough?
To answer this question, we approximate the probability $p(x_{n_c}, n_c)$ by the probability that a noisy trajectory starting from $x_{n_c} \in\TrapInt^{(m,k)}$ is in $\TrapInt$ at $n=n_f$.
This is valid when the noise amplitude $A$ is small enough to ignore the additional probability associated with other basin intervals
for $f_{\mu_f}$ located 
outside $\TrapInt$ (which makes this estimate a slight underestimate).
Let us denote the noiseless trajectory starting from $x_{n_c} \in\TrapInt^{(m,k)}$ by $y_0 (\equiv x_{n_c})$, $y_1, \ldots, y_k$ and the width of the interval $\TrapInt^{(m,k-j)}$, $j=0,1,\ldots,k$ by $2\Delta_j$.
The distribution of the trajectory point $x_{n_c+1}$ at $n=n_c+1$ will be Gaussian with mean $y_1$ and standard deviation $A$.
If we assume that the action of the dynamics on the noise term is linear, the variable $x_{n_c + 2}$ can be written as the sum of two Gaussian variables: a scaled version of the noise from the previous step (mean $y_2$ and standard deviation $A\lambda_2$, where $\lambda_2 \equiv \Delta_2/\Delta_1$ is the expansion factor from $n=n_c+1$ to $n=n_c+2$) and the newly added noise (mean $y_2$ and standard deviation $A$).
Thus, $x_{n_c + 2}$ will be Gaussian with mean $y_2$ and standard deviation $A\sqrt{\lambda_2^2 + 1}$.
Continuing 
in this manner
until the trajectory enters $\TrapInt^{(m)}$ at $n = n_c + k$, we see that the distribution of $x_{n_c + k}$ will be Gaussian with mean $y_k$ and standard deviation $\sigma_k$, where $\sigma_k^2 = A^2(1 + \sum_{i=2}^k L_i^2)$ and $L_i = \prod_{j=i}^k \lambda_j$.
The probability $1-p(x_{n_c}, n_c)$ can then be approximated by the probability that $x_{n_c + k}$ falls outside the interval 
$\TrapInt^{(m)}$. This
leads to the formula
\begin{equation}\label{prob_func_est}
p(x_{n_c}, n_c) = 1 - \frac{\erfc(x_+) + \erfc(x_-)}{2},
\end{equation}
where $\erfc$ denotes the complementary error function, $x_{\pm} = \frac{\Delta_0 \pm x_{n_c}}{A}\cdot f(\lambda_1,\ldots,\lambda_k)$, and $f(\lambda_1,\ldots,\lambda_k)^2 = L_1^2 / (1 + \sum_{i=2}^k L_i^2)$.
Note that it is very unlikely that the trajectory jumps out of the large trapping intervals, $\TrapInt^{(j)}$, $j = m-1,\ldots,0$, when the noise is small, and we have thus assumed that the associated probability is negligible.
Applying this method for approximating $p(x_{n_c}, n_c)$ to the case of Eq.\,\eqref{BNO_map} leads to the theoretical prediction shown in Fig.\ \ref{figure_prob_func}.
We see excellent agreement with numerical simulation within each interval $\TrapInt^{(m,k)}$ for small enough $A$.
For $A = 10^{-4}$, the noise amplitude is larger than the width of all the intervals shown in the figure, and Eq.\ \eqref{prob_func_est} gives an underestimate as expected.
This underestimate, however, does not play a role in
determining the minimum perturbation size because the estimate appears to be accurate whenever $p(x_{n_c}, n_c)$ is large and close to the threshold $\pth$.

In Fig.\ \ref{figure_basin_int}, the box indicating a basin interval $\TrapInt^{(m,k)}$ is colored green if $p(x_{n_c}, n_c)$ estimated by Eq.\ \eqref{prob_func_est} is greater than $\pth$ and red if it is less than $\pth$.
Our argument above predicts that the minimum-size perturbation for effective control at a given time $n_c$ on a trajectory is determined by the closest green box among those associated with $n=n_c$.
This prediction is verified by the result of our numerical experiment shown with blue vertical lines in Fig.\ \ref{figure_basin_int} for a representative reference trajectory.
Indeed, our prediction for the minimum size of control perturbation matches well with the numerical simulation for small enough noise amplitude and up to the time when the trajectories leave the neighborhood of the saddle-node bifurcation point.
This can be seen by comparing the black plots with the corresponding gray plots in Fig.\ \ref{figure_min_kick_size} for $A \le 10^{-4}$ and $n$ up to around $160$.
The rise in the minimum perturbation size after the second minima observed in Fig.\ \ref{figure_min_kick_size} appears to be due to trajectories moving away from the special basin intervals $\TrapInt^{(m,k)}$.

In general, the expansion factors $\lambda_j$ by which noise is amplified along the trajectory are also factors by which the intervals are expanded, and thus determine the interval widths $2\Delta_j$.  It follows from this that $p(x_{n_c}, n_c)>\pth$ if the noise amplitude $A$ is comparable to or smaller than $\Delta_0$, where we recall that $2\Delta_0$ is the width of the basin interval $\TrapInt^{(m,k)}$.
The width of these basin intervals are comparable to the distances between them, and the fractal basin structure guarantees the existence of very narrow basin intervals, typically much narrower than the noise amplitude $A$.
Hence, we expect that the minimum size of perturbation ensuring $p(x_{n_c}, n_c)>\pth$ is comparable to $A$.
Based on this argument, we expect that tipping point control is generally possible with a carefully chosen perturbation of size comparable to noise.

We also expect to observe a ``window of opportunity'' for tipping point control in a wide range of systems. 
To see this, note that immediately after the bifurcation point,
the assumed fractal structure of the basin boundary implies that the basin intervals $\TrapInt^{(m,k)}$ are very small near the bifurcation point.
Hence, at that time, the nearest interval for which $p(x_{n_c}, n_c)>\pth$ tends to be relatively far, making the perturbation required for control relatively large.
Since these basin intervals become larger if we wait longer after the bifurcation, the required perturbation size tends to decrease.
However, if we do not control the trajectory for too long after the bifurcation, and if the trajectory is still in a basin interval of an undesirable attractor (defined analogously to $\TrapInt^{(m,k)}$), then a large perturbation becomes necessary again because 
this undesirable basin interval becomes too large.
We thus expect a short period of time during which the required perturbation size remains small.

\section{Conclusions}

In this paper, we have studied 
tipping point control in
a system undergoing a saddle-node bifurcation with slow variation of its parameter.
When there are multiple attractors after the bifurcation and their basins share a fractal boundary, the arbitrarily fine-scale structure of the interlacing basins allows one to steer an ideal, noise-free trajectory of the system
state
from one basin to another with an arbitrarily small perturbation.
In real systems, however, dynamical noise is unavoidable, and this determines the smallest-scale structure that can be exploited for control.
A careful study of the size of the smallest perturbative control revealed that the best time to apply a control is 
some time after the 
bifurcation takes place.
We have provided a theoretical explanation for this behavior by considering sequences of carefully constructed intervals that comprises the basin of the desirable attractor.
We estimated the probability that a trajectory stays in these intervals until it is captured by a trapping interval for the attractor, and used it to determine an appropriate perturbation for control.
The predictions derived from this calculation 
were
verified to agree with direct numerical simulations within the range of validity of the approximations 
employed.

While the analysis was carried out for one-dimensional systems, we expect that similar behavior would be observed and the tipping point control to be effective also for 
a higher-dimensional system when a
parameter is slowly varied through a saddle-node bifurcation taking place on a fractal basin boundary.
This is because the saddle-node bifurcation is generic in systems of arbitrary dimension and the dynamics of a system at this type of bifurcation can be reduced to that on the associated one-dimensional center manifold\ \cite{center_manifold}.
Starting from the intersection of this center manifold and a trapping region for a desirable attractor, one can construct sequences of segments of the center manifold analogous to the intervals $\TrapInt^{(m,k)}$ discussed above.
It should also be straightforward to extend our approach to continuous-time systems by considering the associated Poincar\'e or stroboscopic maps.

Our results may help design an intervention for controlling a system undergoing a saddle-node bifurcation with multiple post-bifurcation attracting states, some of which may be undesirable.
They indicate that a small intervention may be sufficient, but it is generally better to apply it within a ``window of opportunity.''
Much effort has been made\ \cite{Scheffer:2009fk} to determine whether and when the system is approaching such a bifurcation point.
We suggest that knowing the bifurcation point is not the entire story --- how to design  
an effective, low-amplitude intervention may depend non-trivially on the properties of the bifurcation as well as on external noise.
Given that the saddle-node bifurcation is generic and that fractal basin boundaries are common, we expect that our results will stimulate further investigation and encourage application to real systems.

\begin{acknowledgments}
This work was supported by MURI contract ONR N00014-07-1-0734.
\end{acknowledgments}

%

\end{document}